\documentclass[conference]{IEEEtran}
\IEEEoverridecommandlockouts
\usepackage{cite}
\usepackage{amsmath,amssymb,amsfonts}
\usepackage{algorithmic}
\usepackage{graphicx}
\usepackage{textcomp}
\usepackage{xcolor}
\usepackage{balance}
\usepackage{todonotes}

\newcommand{\etal}{\textit{et al.~}}

\def\BibTeX{{\rm B\kern-.05em{\sc i\kern-.025em b}\kern-.08em
    T\kern-.1667em\lower.7ex\hbox{E}\kern-.125emX}}

\begin{document}

\title{Cross-Representation Transferability of Adversarial Attacks: From Spectrograms to Audio Waveforms}
\author{\IEEEauthorblockN{Karl Michel Koerich\IEEEauthorrefmark{1}, Mohammad Esmaeilpour\IEEEauthorrefmark{2},
Sajjad Abdoli\IEEEauthorrefmark{2},
Alceu de S. Britto Jr.\IEEEauthorrefmark{3}\\ and Alessandro L. Koerich\IEEEauthorrefmark{2}}\\
\IEEEauthorblockA{\IEEEauthorrefmark{1}McGill University, Montreal, QC, Canada\\ Email: karl.koerich@mail.mcgill.ca}
\IEEEauthorblockA{\IEEEauthorrefmark{2}\'{E}cole de Technologie Superi\'{e}ure, Universit\'{e} du Qu\'{e}bec, Montreal, QC, Canada\\
Email: mohammad.esmaeilpour.1@ens.etsmtl.ca, sajjad.abdoli.1@ens.etsmtl.ca, alessandro.koerich@etsmtl.ca}
\IEEEauthorblockA{\IEEEauthorrefmark{3}Pontifical Catholic University of Paran\'{a}, Curitiba, PR, Brazil\\
Email: alceu@ppgia.pucpr.br}
}

\maketitle

\begin{abstract}
This paper shows the susceptibility of spectrogram-based audio classifiers to adversarial attacks and the transferability of such attacks to audio waveforms. Some commonly used adversarial attacks to images have been applied to Mel-frequency and short-time Fourier transform spectrograms, and such perturbed spectrograms are able to fool a 2D convolutional neural network (CNN). Such attacks produce perturbed spectrograms that are visually imperceptible by humans. Furthermore, the audio waveforms reconstructed from the perturbed spectrograms are also able to fool a 1D CNN trained on the original audio. Experimental results on a dataset of western music have shown that the 2D CNN achieves up to 81.87\% of mean accuracy on legitimate examples and such performance drops to 12.09\% on adversarial examples. Likewise, the 1D CNN achieves up to 78.29\% of mean accuracy on original audio samples and such performance drops to 27.91\% on adversarial audio waveforms reconstructed from the perturbed spectrograms.
\end{abstract}


\section{Introduction}
\label{sec:intro}
Music genre classification is a challenging task for humans \cite{costa2004automatic,silla2008machine,lidy2010suitability, Costa2011,Costa2012a,Koerich13} due to the subjectivity and unclear boundaries between genres, and the uniqueness of musicians and artists. Yet, well classifying music is of great interest to many researchers and companies in the entertainment and arts industry. In the last years, convolutional neural networks (CNNs) became increasingly popular due to their high accuracy and performance on image datasets. Therefore, the focus in academia has been on 2D CNNs. When it comes to audio and music processing, CNNs have had a significant impact on several tasks such as automatic music tagging \cite{dieleman2014end}, video clip classification based on audio information \cite{hershey2017cnn}, speaker identification \cite{Ravanelli2018}, environmental sound classification \cite{esmailpour2019,abdoliCK19} and music genre classification \cite{Pons2018,Choi2017,kereliuk2015deep}.

Even if audio is a 1D signal, it is a common practice to use 2D representations, like spectrograms, when training machine learning models. Due to their ability to model the human peripheral auditory system \cite{tiwari}, Mel-frequency cepstrum coeficient (MFCC) features are currently used for several audio processing tasks such as music genre classification \cite{dan2017}. Mel-frequency (MF) spectrograms, which use Mel-frequency filter banks to represent the short-term power spectrum of audio on the Mel-scale of frequency, are one of the most preferred input types for music information retrieval \cite{texas}. One of the main advantages of using 2D representations is that they summarize high dimensional waveforms into compact time-frequency representations while audio signals alone are noisier \cite{stowell2014}. Regardless of the type of spectrogram; as they are 2D representations of audio signals, they can be treated as images. Therefore, this opens up the opportunity to benefit from the recent advances of deep neural networks in computer vision.

Recent works have exploited the capability of CNNs to learn representations directly from spectrograms. Boddapati \etal\cite{boddapati2017classifying} use short-time Fourier transform (STFT), MFCC and Cross Recurrence Plot (CRP) spectrograms with two different 2D CNN architectures (AlexNet and GoogLeNet) to classify 2D representations of environmental sounds. Lee \etal\cite{Lee:Nam} use 2D CNNs with MF spectrograms as input for music auto-tagging. Pons \etal\cite{Pons2016} use 40 bands MF spectrograms to experiment with musically motivated CNNs and try understanding what CNNs learn from particular datasets. Pons \etal\cite{Pons2018} use MF spectrograms as input for a randomly weighted CNN for music audio classification. Oramas \etal\cite{Oramas2017} use constant-Q transform (CQT) spectrograms in their audio-based approach for multi-label music genre classification.

Despite all advantages, it has been shown that approaches based on 2D representations are susceptible to adversarial attacks, which can easily fool these models and raise safety concerns. Esmailpour \etal\cite{esmaeilpour2019robust} have shown that the majority of state-of-the-art approaches for audio classification relying on 2D CNNs can be easily deceived, with fooling rates higher than 90\% and high confidence. 1D CNNs can also be easily fooled by adversarial attacks. Abdoli \etal\cite{abdoli2019} demonstrated the existence of universal adversarial perturbations that can fool several audio processing architectures with attack success rates between 91.1\% and 74.7\%, and signal-to-noise ratio (SNR) between 15.70dB and 19.62dB. Du \etal\cite{du2019sirenattack} proposed a method based on Particle Swarm Optimization (PSO) for generating adversarial audio for end-to-end audio systems. They evaluated their attacks on a range of applications like speech command recognition, speaker recognition, sound event detection and music genre classification. The proposed attack achieved a success rate of 89.30\% on a 1D CNN and 91.20\% on a convolutional recurrent neural network with SNR of 15.39dB and 17.24dB respectively. However, the low SNRs indicate that the adversarial perturbations are audible and can be easily perceived by a listener.

{\color{black} Even if few works have already studied adversarial attacks on both 1D and 2D CNNs \cite{du2019sirenattack, abdoli2019, kereliuk2015deep, esmaeilpour2019robust}, none of them have evaluated the transferability of such adversarial attacks across representations, in particular from 2D to 1D. Generating 1D adversarial attacks is much more time-consuming than generating 2D attacks due to the high dimensionality of audio signals \cite{esmaeilpour2019robust}, as it requires computing a similarity measure such as the $\ell_2$-norm between legitimate and crafted examples as a part of an adversarial optimization criterion. Therefore, it might be advantageous to generate 2D perturbations and transfer it to audio waveforms.}

The main contributions of this paper are: {\color{black} (i) we show that the most effective adversarial attacks for images can also attack spectrograms generated from music, which reveals the vulnerability of 2D CNNs to non-specific attacks; (ii) we show that perturbed spectrograms can be used to reconstruct audio signals that are perceptually similar to the original audio, with an SNR value of about 20dB; (iii) we show that the audio waveforms reconstructed from the perturbed spectrograms can also fool a 1D CNN trained on the original audio with high confidence.}

This paper is organized as follows. Section \ref{sec:arch} presents the baseline 1D and 2D CNN architectures for music genre classification. Section \ref{sec:adv} presents a description of the adversarial attacks and the reconstruction process of the audio waveforms from spectrograms. Section \ref{sec:exp} presents the dataset used, the experimental protocol, and the experimental results. We compare the performance of the 2D CNN model when using MF and STFT spectrograms, and the vulnerability of such a model to adversarial examples. {\color{black} We also evaluate the transferability of adversarial perturbations from 2D representations to the audio waveform and the susceptibility of a 1D architecture to such a transferred attack.} Finally, the conclusions and perspectives of future work are presented in the last section.

\section{1D and 2D CNN Architectures}
\label{sec:arch}
\begin{figure}[htpb!]
  \centering
  \includegraphics[width=0.48\textwidth]{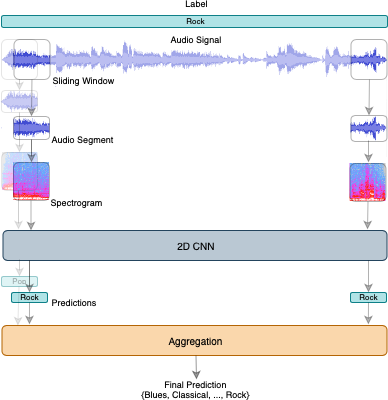}
  \caption{Overview of the proposed approach for audio signal segmentation by a sliding window, transformation of audio segments into spectrograms, prediction by a 2D CNN and aggregation of prediction on the segments to come up to a final prediction.}
  \label{fig:over}
\end{figure}

\begin{figure}[htpb!]
  \centering
  \includegraphics[width=0.48\textwidth]{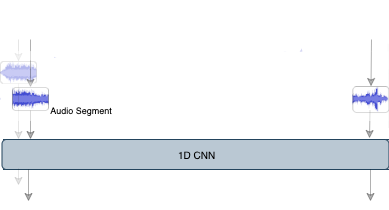}
  \caption{Overview of the changes in Fig.~\ref{fig:over} for the 1D approach. The 1D CNN receives the waveform segments instead of spectrograms and provides the predictions on segments for aggregation.}
  \label{fig:over1D}
\end{figure}

The aim of the proposed architecture is to deal with 2D representations of audio signals of various lengths, learning meaningful representations directly from spectrograms. First, we split each audio waveform into fixed-length segments using a sliding window of appropriate width. The window width depends mainly on the signal sampling rate, which in the case of the music dataset evaluated in this paper is 22,050 Hz. In our approach, we use a window of five seconds (110,250 samples) because such a length provides the best trade-off between the number of segments and accuracy. Furthermore, there is also a certain percentage of overlapping between successive audio segments, which aim is to maximize the use of information. In our approach we use 75\% overlapping because such percentage of overlap provides the best trade-off between the number of segments and accuracy. Furthermore, the overlapping can be viewed as some sort of data augmentation since it naturally increases the number of samples due to the fact that some parts of the audio signal are reused. Fig.~\ref{fig:over} summarizes the process of splitting the audio file into appropriate segments by the sliding window, then having the signals converted to spectrograms, which are then used as input to the 2D CNN. A similar process is used with the 1D CNN except that the input of the CNN is audio segments, as illustrated in  Fig.~\ref{fig:over1D}.

Moreover, the length of the original audio (before being split) has a direct impact on the number of samples being tested and trained, which may impact the computational cost of the model. The GTZAN dataset has a sampling rate of 22,050 Hz and all original audio files are 30 seconds long.

STFT and MF are the main approaches for producing spectrograms for music signals. To generate the STFT and MF spectrograms, we use a fast Fourier transform (FFT) window of length 512 and 256 samples between successive frames. For the MF spectrograms we use 64 Mel filters. Therefore, the spectrogram dimension becomes 431$\times$257 for STFT, and 431$\times$64 for MF. {\color{black} The architecture of the 2D CNN is the same for both input formats and it has two convolutional layers (CL), followed by a maxpooling layer to reduce the complexity of the network, followed by another two CLs and another maxpooling layer, which is connected to a fully connected (FC) layer followed by an output layer. ReLU is used as activation function in all layers but the last which uses softmax. The weights of all of layers are initialized randomly. Batch normalization is included between CLs and a dropout of 0.4 is used after the FC layer. The breakdown of the 2D CNN is presented in Table~\ref{tab:arch}.}

{\color{black} 
Several 1D CNN architectures have been proposed to deal directly on the audio waveforms \cite{abdoli2019}. The 1D CNN model receives 5-second audio segments as input and it has five one-dimensional CLs, where the first CL employs a Gammatone filter-bank. This layer is kept frozen during the training process. Gammatone filters are used to decompose the input audio signal to appropriate frequency bands. The output of the last CL is used as input to one FC layer followed by an output layer. Leaky ReLU is used as activation function in all layers except in the last which uses softmax. The weights of all of the layers are initialized randomly. This model was proposed by Abdoli \etal\cite{abdoliCK19} for environmental sound classification. Batch normalization is included between CLs and a dropout of 0.4 is used after the FC layer. The breakdown of the 1D CNN is presented in Table~\ref{tab:arch1D}.}

\begin{table}[htpb!]
\centering
\footnotesize
\renewcommand{\arraystretch}{1.3}
\caption{Architecture of the proposed 2D CNN}
\begin{tabular}{|l|l|l|l|l|l|}
\hline
\multicolumn{1}{|l|}{\textbf{Layer}}&\bf \# of & \bf Filter & \bf Stride &\multicolumn{2}{|c|}{\textbf{Output Shape}} \\
\textbf{Type}             & \bf Filters& \bf Size & &\multicolumn{1}{c}{\bf STFT}&\multicolumn{1}{c|}{\bf MF}\\
\hline
\multicolumn{1}{|l|}{Input}                 &-&-&-& 257, 431, 1 & 64, 431, 1            \\
\hline
\multicolumn{1}{|l|}{Conv2D}                &32&3, 3&1, 1& 255, 429, 32 &62, 429, 32          \\
\hline
\multicolumn{1}{|l|}{Conv2D}                &32&3, 3&1, 1& 253, 427, 32 &60, 427, 32           \\
\hline
\multicolumn{1}{|l|}{MaxPool}        &-&-&2, 2& 126, 213, 32 &30, 213, 32          \\
\hline
\multicolumn{1}{|l|}{Conv2D}                &64&3, 3&1, 1& 124, 211, 64 &28, 211, 64      \\
\hline
\multicolumn{1}{|l|}{Conv2D}                &64&3, 3&1, 1& 122, 209, 64 &26, 209, 64         \\
\hline
\multicolumn{1}{|l|}{MaxPool}        &-&-&2, 2& 60, 103, 64  &12, 103, 64             \\ 
\hline
\multicolumn{1}{|l|}{Dense}                 &-&-&-        & 1~024         &1~024   \\
\hline
\multicolumn{1}{|l|}{Output}                 &-&-&-        & 10           &10      \\
\hline
\end{tabular}
\label{tab:arch} 
\end{table}

\begin{table}[htpb!]
\centering
\footnotesize
\renewcommand{\arraystretch}{1.2}
\caption{Architecture of the proposed 1D CNN}
\begin{tabular}{|l|l|l|l|l|}
\hline
\multicolumn{1}{|l|}{\textbf{Layer}}&\bf \# of & \bf Filter & \bf Stride &\multicolumn{1}{|c|}{\textbf{Output}} \\
\textbf{Type}             & \bf Filters& \bf Size & &\multicolumn{1}{c|}{\bf Shape}\\
\hline
\multicolumn{1}{|l|}{Input}                 &-&-&-& 110~250            \\
\hline
\multicolumn{1}{|l|}{Conv1D Gamma}                & 32 & 512 & 1 & 109~739          \\
\hline
\multicolumn{1}{|l|}{AvgPool}        &-&-&8& 13~717          \\
\hline
\multicolumn{1}{|l|}{Conv1D}                & 16 & 256 & 2 & 6~731            \\
\hline
\multicolumn{1}{|l|}{AvgPool}        &-&-&8& 841        \\
\hline
\multicolumn{1}{|l|}{Conv1D}                & 32 & 64 & 2 & 389       \\
\hline
\multicolumn{1}{|l|}{Conv1D}                & 64 & 32 & 2 & 179          \\
\hline
\multicolumn{1}{|l|}{Conv1D}                & 128 & 16 & 2 & 82         \\
\hline
\multicolumn{1}{|l|}{MaxPool}        & - & - & 2 & 41              \\ 
\hline
\multicolumn{1}{|l|}{Dense}                 &-&-&-        & 256            \\
\hline
\multicolumn{1}{|l|}{Output}                 &-&-&-        & 10                \\
\hline
\end{tabular}
\label{tab:arch1D} 
\end{table}


\textcolor{black}{During the classification step, since the input audio waveform is split into several segments, we need to aggregate the 2D CNN predictions to come up to a final decision on the input audio, as illustrated in Fig.~\ref{fig:over}. For such an aim, we used majority vote and the sum rule \cite{abdoliCK19}. When there are $K$ classes, we generate $K$ values and we choose the class which has the maximum value as our final prediction. A similar process is used to aggregate the predictions of the 1D CNN, as illustrated in Fig.~\ref{fig:over1D}.}

\section{Adversarial Attacks and Audio Reconstruction}
\label{sec:adv}
Adversarial attacks can be considered as small crafted perturbations that, when intentionally added to a legitimate example, lead machine learning models to misbehave \cite{Weng2018}. Considering $x$ as a legitimate example, then an adversarial example $x^\prime$ can be crafted in such a way that:

\begin{equation}
x \approx x^\prime, \qquad f^{*}(x)\neq f^{*}(x^\prime)    
\end{equation}

\noindent where $f^{*}$ is the post-activation function. Assuming that $x$ is a spectrogram, the crafted $x^\prime$ should be {\color{black} unrecognizable by human visual system.}


\begin{figure}[htpb!]
  \centering
  \includegraphics[width=0.5\textwidth]{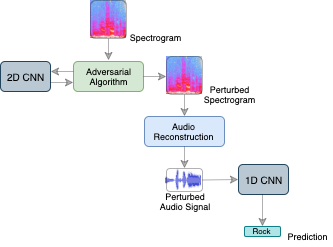}
  \caption{Overview of the 2D adversarial attack which relies on the 2D CNN model and training dataset (spectrograms) to produce adversarial spectrograms. The audio waveform is reconstructed from the perturbed spectrogram and used to fool the 1D CNN model.}
  \label{fig:advattack}
\end{figure}

Among the several algorithms for generating $x^\prime$, the Fast Gradient Sign Method (FGSM) \cite{fgsm} was one of the first attacks, which still remains one of the most effective adversarial attacks. Goodfellow \etal\cite{fgsm} introduced the FGSM for generating simple adversarial samples. {\color{black} The method consists of adding to the legitimate example $\bold{x}$ an imperceptibly small perturbation that is equal to the product of a small constant $\epsilon$ and the sign of the gradient of the cost function $J$ for the model parameter $\bold{w}$ with respect to the input $\bold{x}$ and the true label $y$, as shown in \eqref{eq:FGSM}.

\begin{equation}
\bold{x}^\prime = \bold{x}+\epsilon \cdot \mathrm{sign}(\nabla_{\bold{x}}J(\bold{w}, \bold{x}, y))
\label{eq:FGSM}
\end{equation}

\noindent The resulting adversarial example $\bold{x}^\prime$ carries a small perturbation that cannot be seen by the human eye and effortlessly deceives 2D CNNs and other non-deep architectures \cite{esmaeilpour2019robust}.}

Kurakin \etal\cite{kurakin2016adversarial} introduced a straightforward way of extending the FGSM method by applying it multiple times with a small step size. Known as the Basic Iterative Method (BIM), this adversarial attack is also able to fool complex 2D CNNs. As illustrated in the upper part of Fig.~\ref{fig:advattack}, both FGSM and BIM are white-box adversarial attacks which means that both the trained 2D CNN model and the training dataset should be accessible to the adversarial algorithms to fetch its gradient information and generate the adversarial input $\bold{x}^\prime$ with unrecognizable differences to the legitimate input $\bold{x}$. The perturbed spectrogram can perhaps make a 2D CNN predict a wrong label with high confidence. {\color{black} The lack of robustness of 2D CNNs to these two attacks was also observed for the task of environmental sound classification \cite{esmaeilpour2019robust,esmailpour2020detection}}.


\begin{figure}[htpb!]
  \centering
  \includegraphics[width=0.49\textwidth]{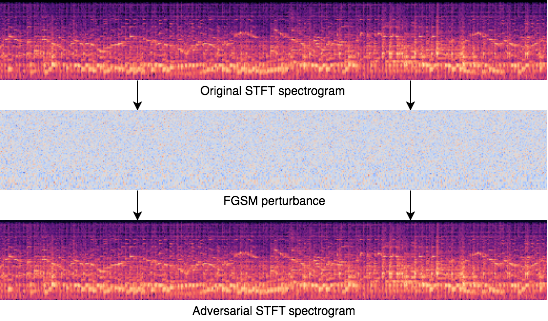}
  \caption{The original STFT spectrogram and the perturbation produced by the FGSM attack to generate an adversarial STFT spectrogram.}
  \label{fig:fgsm}
\end{figure}

Fig.~\ref{fig:fgsm} shows an example of a legitimate STFT spectrogram attacked with a perturbation produced by the FGSM attack. {\color{black}Such a figure summarizes what happens with attacked STFT and MF spectrograms: the difference between the adversarial and legitimate spectrograms is imperceptible to the human visual system. {\color{black} This aspect is very important, otherwise the perturbation could not be considered as an adversarial one.}} However, what happens with the perturbed signal in its original representation space, that is, audio waveform? Will such a perturbation remain unperceived by the human auditory system?


\subsection{Audio Reconstruction}
The main contribution of this paper is to evaluate if after performing adversarial attacks to the 2D representations and successfully fooling the 2D CNN, such attacks could be transferred to the audio waveform. If so, the perturbations should not be perceptible by human auditory system on the audio after reconstructing it from the perturbed spectrogram. {\color{black} Besides evaluating the transferability of the adversarial attacks across representations (from 2D to 1D), we also want to evaluate if such an reconstructed adversarial audio is able to fool an 1D CNN model which accomplishes the same task of the 2D CNN. For such an aim, we need to reconstruct the audio signals from the spectrograms.}

{\color{black} The phase of the audio signal is fundamental to an accurate reconstruction of the signal. However, both the STFT and MF spectrograms do not contain information about the exact, or even approximate, phase of the signal that it represents. Therefore, it is not possible to reverse the process without distorting the reconstructed signal.

To circumvent this problem, we can retain the phase information separately and, depending on the type of spectrogram, we can use such information to reconstruct the original audio signal, which is the case of the STFT spectrogram. This avoids the introduction of distortions to the reconstructed audio signal, which may tamper or even mask the adversarial perturbations embedded into the reconstructed adversarial audio examples. On the other hand, to reconstruct the audio signals from the MF spectrograms, it is necessary to estimate the unknown phases iteratively using the Griffin-Lim (GL) algorithm \cite{griffin84}.}

\section{Experimental Results}
\label{sec:exp}
The proposed 2D CNN for music genre classification was evaluated on the GTZAN dataset. This dataset consists of 1,000 30-second audio clips evenly distributed into 10 classes: Blues, Classical, Country, Disco, Hip-Hop, Jazz, Metal, Pop, Reggae and Rock. The audio samples were collected from a variety of sources in order to represent a variety of recording conditions \cite{Tzanetakis:2002}. {\color{black} Even if the GTZAN dataset has several known problems with its integrity, such as replications, mislabeling, and distortions \cite{Sturm2012, kereliuk2015deep}, this does not affect our experiments since our aim is not assessing the accuracy of CNN models, but their vulnerabilities against adversarial attacks.}

The 1,000 audio files were shuffled and divided into three folds with 340, 330 and 330 samples, respectively. Fold 1 contains 34 tracks of each of the 10 genres; Folds 2 and 3 contain 33 tracks of each genre. Every track of every fold was split into 21 short segments according to the sliding window described in Fig.~\ref{fig:over}. The next step was to generate spectrograms of all 21,000 audio segments. The model was first accessed with MF spectrograms and then STFT spectrograms. Two folds were used for training and 20\% of the training set was used for validation. The third fold was used for testing. After predicting the music genre for each segment on the testing set, the predictions for all 21 windows belonging to the same song are aggregated with majority vote and sum rule to determine the final genre prediction for the whole track. Each network was trained up to 100 epochs with batch sizes of 50 samples and early stopping. The model's performance for each input type is presented in Table \ref{table:acc_models}. {\color{black} The performance reported in Table~\ref{table:acc_models} is far from the best performance already achieved for such a dataset which is about 90\% \cite{PanagakisKA09}. On the other hand, the results are similar to Kereliuk \etal\cite{kereliuk2015deep} that used two different 2D CNN architectures and achieved 81.20\% of mean accuracy for the same dataset. Furthermore, we did not attempt to fully optimize the performance of the 2D CNN to reduce the risk of overfitting, as it can potentially increase the susceptibility of CNNs to adversarial attacks.}

\begin{table}[htpb!]
\caption{Mean accuracy and standard deviation of the 2D CNN for STFT and MF spectrograms.}
\centering
\renewcommand{\arraystretch}{1.3}
\begin{tabular}{|l|c|c|c|}
\hline
 & \multicolumn{3}{|c|}{\textbf{Mean Accuracy $\pm$ SD (\%)}} \\ 
\cline{2-4}
 \textbf{Input}            & \multicolumn{1}{|c|}{\textbf{Segments}}           &\multicolumn{1}{|c|}{\textbf{MV Aggregation}}        &\multicolumn{1}{|c|}{\textbf{SR Aggregation}}   \\ 
\hline
STFT                & {67.09$\pm$1.49} &{75.84$\pm$1.92}  &{75.64$\pm$2.09}\\
 MF                  & {75.29$\pm$2.50} &{81.87$\pm$2.49}  &{81.37$\pm$2.20}\\
\hline
\multicolumn{4}{c}{\scriptsize{MV: Majority Voting; SR: Sum Rule; SD: Standard Deviation }}
\end{tabular}
\label{table:acc_models}
\end{table}

{\color{black} Table~\ref{table:acc_models1D} shows the performance achieved by the 1D CNN on original audio waveforms as well as on the waveforms reconstructed from the STFT spectrogram with the original phase information. The performance achieved on the reconstructed audio is slight better than that achieved on the original audio. This is a clear indication that the reconstruction process is accurate and it does not insert spurious noise to the signal.}

\begin{table}[htpb!]
\caption{Mean accuracy and standard deviation of the 1D CNN for original audio and audio signal reconstructed from STFT spectrogram and phase information.}
\centering
\renewcommand{\arraystretch}{1.3}
\begin{tabular}{|l|c|c|c|}
\hline
 & \multicolumn{3}{|c|}{\textbf{Mean Accuracy $\pm$ SD (\%)}} \\ 
\cline{2-4}
 \textbf{Input}            & \multicolumn{1}{|c|}{\textbf{Segments}}           &\multicolumn{1}{|c|}{\textbf{MV Aggregation}}        &\multicolumn{1}{|c|}{\textbf{SR Aggregation}}   \\ 
\hline
Original                 & {70.82$\pm$0.20} &{77.69$\pm$0.81}  &{77.99$\pm$1.04}\\
STFT Recons                   & {71.87$\pm$0.67} &{78.49$\pm$0.26}  &{78.29$\pm$0.72}\\
\hline
\multicolumn{4}{c}{\scriptsize{MV: Majority Voting; SR: Sum Rule; SD: Standard Deviation }}
\end{tabular}
\label{table:acc_models1D}
\end{table}

Table~\ref{table:fgsm} shows the results of the FGSM and BIM attacks against the 2D CNN. We evaluate the mean accuracy achieved by the model on the perturbed spectrograms. For both STFT and MF spectrograms the BIM attack is more successful due to its iterative nature. For instance, considering the best result of Table~\ref{table:acc_models}, the mean accuracy for STFT spectrogram drops from 75.84\% to 11.58\% and from 81.87\% to 12.09\% for the MF spectrogram.  

\begin{table}[htpb!]
\caption{Mean accuracy and standard deviation for the 2D CNN after attacking STFT and MF spectrograms with FGSM and BIM adversarial attacks.}
\centering
\renewcommand{\arraystretch}{1.3}
\begin{tabular}{|l|c|c|c|c|}
\hline
    &  & \multicolumn{3}{|c|}{\textbf{Mean Accuracy $\pm$ SD (\%)}} \\
    \cline{3-5}
\textbf{Input}  & \bf Attack &\multicolumn{1}{|c|}{\textbf{Segments}}           &\multicolumn{1}{|c|}{\textbf{MV Aggreg.}}        &\multicolumn{1}{|c|}{\textbf{SR Aggreg.}}   \\ 
\hline
STFT                & FGSM & {17.40$\pm$1.20} &{17.35$\pm$0.97}  &{16.76$\pm$1.19}\\
                & BIM & {13.26$\pm$1.20} &{11.58$\pm$0.97}  &{\hspace{4pt}9.89$\pm$1.19}\\
\hline
MF                  & FGSM & {21.52$\pm$1.80} &{19.32$\pm$1.46}  &{19.41$\pm$0.88}\\
                  & BIM & {15.38$\pm$1.99} &{12.09$\pm$2.01}  &{12.49$\pm$1.88}\\
\hline
\multicolumn{5}{c}{\scriptsize{MV: Majority Voting; SR: Sum Rule; SD: Standard Deviation }}
\end{tabular}
\label{table:fgsm}
\end{table}

{\color{black} Table~\ref{table:fgsm1D} shows the performance of the 1D CNN on the adversarial audio samples generated by the reconstruction of the STFT spectrogram perturbed with FGSM and BIM. The mean accuracy drops from 77.99\% on the original audio to 27.91\% on the examples attacked by FGSM. We did not evaluate audio examples reconstructed from MF spectrograms because even the reconstruction of legitimate examples is very noisy due to the approximate phase estimation by the GL algorithm.}
 
\begin{table}[htpb!]
\caption{Mean accuracy and standard deviation for the 1D CNN after reconstructing audio from STFT spectrograms attacked with FGSM and BIM adversarial attacks.}
\centering
\renewcommand{\arraystretch}{1.3}
\footnotesize
\begin{tabular}{|l|c|c|c|c|}
\hline
    &  & \multicolumn{3}{|c|}{\textbf{Mean Accuracy $\pm$ SD (\%)}} \\
    \cline{3-5}
\textbf{Input}  & \bf Attack &\multicolumn{1}{|c|}{\textbf{Segments}}           &\multicolumn{1}{|c|}{\textbf{MV Aggreg.}}        &\multicolumn{1}{|c|}{\textbf{SR Aggreg.}}   \\ 
\hline
STFT Recons               & FGSM & {26.45$\pm$1.09} &{28.00$\pm$0.98}  &{27.91$\pm$0.85}\\
                & BIM & {30.76$\pm$2.03} &{33.85$\pm$3.45}  &{33.35$\pm$3.67}\\
\hline
\multicolumn{5}{c}{\scriptsize{MV: Majority Voting; SR: Sum Rule; SD: Standard Deviation }}
\end{tabular}
\label{table:fgsm1D}
\end{table}

{\color{black} Finally, we need to evaluate if the adversarial perturbation added to the spectrogram will remain unrecognizable by human auditory system when the audio waveform is reconstructed from the adversarial spectrogram. For such an aim we conducted two experiments: (i) a quantitative experiment using signal-to-noise ratio (SNR) as a metric to measure the level of the noise with respect to the original signal; (ii) a qualitative listening experiment with expert and non-expert listeners.}

SNR has been used by previous works to evaluate the quality of the generated adversarial audio by measuring the level of the perturbation on the signal after adding the perturbations \cite{abdoli2019, kereliuk2015deep, du2019sirenattack} and it is defined as:

\begin{equation}
\textrm{SNR}_\textrm{dB}({\mathbf x^r},{\mathbf v})=20.\log_{10}{ \frac{P({\mathbf x^r})}{P(\mathbf v)}},
\end{equation}

\noindent where $\mathbf x^r$ denotes the audio reconstructed from the legitimate spectrogram and $\mathbf v$ denotes the adversarial noise. \(P(.)\) is the power of the signal or noise, which is defined as:

\begin{equation}
\label{eq:power}
{P}( {\mathbf x})=\sqrt{\frac{1}{N} \sum_{n=1}^{N}  x_{n}^{2}},
\end{equation}

\noindent where $x_{n}$ denotes the $n$-th component of the signal ${\mathbf x}$. A high SNR indicates that a low level of noise is added to the audio signal by the adversarial perturbation.

According to Du \etal\cite{du2019sirenattack}, the noise is imperceptible when $\textrm{SNR}_\textrm{dB}$ is equal or greater than 20dB. This is also supported by the experiments carried out by Abdoli \etal\cite{abdoliCK19} for environmental sound classification\footnote{See: https://sajabdoli.netlify.com/publication/uap/ for some audio samples}. Table~\ref{tab:snr} shows the mean SNR achieved on the reconstructed audio from the spectrograms attacked by FGSM and BIM, and None refers to the SNR between the original and reconstructed audio. {\color{black} In the case of STFT spectrograms, which uses the phase information in the reconstruction process, the reconstructed audio is more accurate and it becomes equivalent to the original audio (SNR $>$ 90dB). On the other hand, MF spectrograms rely on the phase estimation using Griffin-Lim's method, thus the reconstructed audio is quite noisy (SNR $<$ 20dB). Therefore, the SNR for the audio reconstructed from MF spectrograms is quite misleading because the noisy reconstruction also masks the adversarial perturbation}.

\begin{table}[htpb!]
\label{table:new_snr}
\caption{{\color{black}Mean SNR and standard deviation (SD) computed by \eqref{eq:power} for the audio reconstructed from the perturbed STFT and MF spectrograms by the FGSM and BIM attacks.}}
\centering
\footnotesize
\renewcommand{\arraystretch}{1.3}
\begin{tabular}{|c|c|c|}
\hline 
\textbf{Input}    & \bf Attack & \multicolumn{1}{|c|}{\textbf{SNR$_\textrm{dB}$($\mathbf x, \mathbf v$) $\pm$ SD}} \\
\hline
Audio Reconstructed                            & None & {$>$90}\\
(STFT) & FGSM & {14.62$\pm$5.93}\\
                             & BIM  & {20.19$\pm$5.95}\\
\hline
Audio Reconstructed  & None  & {8.71$\pm$11.77}\\
        (MF)             & FGSM  & {42.69$\pm$9.45}\\    
        & BIM   & {44.21$\pm$9.90}\\
\hline
\end{tabular}
\label{tab:snr}
\end{table}

{\color{black}The qualitative experiment was very limited and it was conducted with only four listeners using the speakers of a desktop computer. Pairs of legitimate and adversarial audio of the same song in a random order were presented to listeners and they pointed out whether they perceive or not differences in the audio samples.}
{\color{black}The results are disappointing because in average, listeners have noticed audible difference in all audio pairs. Nevertheless, most of the listeners referred to a "small background noise" which is not even close to change the perception of the listeners about the musical genre. Several examples of legitimate and adversarial audio samples are available to readers\footnote{https://github.com/karlmiko/ijcnn2020}}.

\section{Conclusion}
\label{sec:concl}

\textcolor{black}{This paper presented a 2D CNN for music genre classification and evaluated it with two 2D representations: MF spectrograms and STFT spectrograms. The proposed approach learns from spectrograms of audio segments and performs relatively well compared to the state-of-the-art on the GTZAN dataset. The proposed 2D CNN achieved 81.87\% and 75.84\% of mean accuracy for MF and STFT spectrograms, respectively. Even if spectrograms seem to be advantageous to model in a compact and informative way the spectrum of frequencies of an audio signal as it varies with time, such 2D representations and 2D CNN models may not be the safest ones when it comes to robustness against adversarial attacks.}

\textcolor{black}{In this paper we have shown that adversarial attacks designed to regular images can also harm 2D representation of audio signals since the perturbations remain visually imperceptible on spectrogram images. FGSM and BIM attacks successfully fooled the 2D CNN in the task of music genre classification. These adversarial examples produced using 2D representation and a 2D CNN model were then transferred to audio waveforms. The audio waveforms produced were tested against a 1D CNN model and successfully fooled it. Therefore, we have shown the transferability of adversarial perturbations across representations. These results expose the vulnerability of both 1D and 2D CNN architectures to adversarial attacks.}

\textcolor{black}{The audio signals reconstructed from STFT spectrograms using the phase information have a very high SNR and the adversarial audio reconstructed from such spectrograms have SNRs between 14dB and 20dB. Nevertheless, the reconstructed adversarial audio examples are distinguishable from their legitimate counterparts to human perception according to the outcome of the qualitative evaluation. Even if the audio signals reconstructed from MF spectrograms do not have SNRs as high as those achieved using STFT spectrograms, the adversarial audio reconstructed from MF spectrograms have SNR greater than 20dB, which is mostly due to the noise introduced in the reconstruction process and not the adversarial perturbations.}

\textcolor{black}{Future work related to this paper includes adding acoustic perceptual considerations to eliminate spectral components in a way that no difference is perceived by the listeners. Also, since we showed the transferability of adversarial perturbations, future work includes the possibility of creating black-box adversarial attacks to 1D CNN models by using auxiliary 2D CNN models to produce adversarial images and then have them transferred to audio waveforms.}

\balance

\bibliographystyle{IEEEbib}


\end{document}